# Modulation of field-like spin orbit torque in heavy metal / ferromagnet heterostructure

Zilu Wang[1], Houyi Cheng[1,3], Kewen Shi[1,*], Yang Liu[1], Junfeng Qiao[1], Daoqian Zhu[1], Wenlong Cai[1], Xueying Zhang[1,2], Sylvain Eimer[1,3], Dapeng Zhu[1,2], Jie Zhang[1], Albert Fert[1,4] and Weisheng Zhao[1,2,3,†]

[1]*Fert Beijing Institute, BDBC, School of Microelectronics, Beihang University, Beijing, China.*
[2]*Beihang-Goertek Joint Microelectronics Institute, Qingdao Research Institute, Beihang University, Qingdao, China.*
[3]*Hefei Innovation Research Institute, Beihang University, Hefei 230013, China*
[4]*Unité Mixte de Physique, CNRS, Thales, University of Paris-Saclay, Palaiseau, France.*

*shikewen@buaa.edu.cn; †weisheng.zhao@buaa.edu.cn

# ABSTRACT

Recent studies rediscovered the crucial role of field-like spin orbit torque (SOT) in nanosecond-timescale SOT dynamics. However, there is not yet an effective way to control its relative amplitude. Here, we experimentally modulate the field-like SOT in W/CoFeB/MgO trilayers through tuning the interfacial spin accumulation. By performing spin Hall magnetoresistance measurement, we find that the CoFeB with enhanced spin dephasing, either generated from larger layer thickness or from proper annealing, can distinctly boost the spin absorption and enhance the interfacial spin mixing conductance $G_r$. While the damping-like torque efficiency increases with $G_r$, the field-like torque efficiency turns out to decrease with it. The results suggest that the interfacial spin accumulation, which largely contributes to a field-like torque, is reduced by higher interfacial spin transparency. Our work shows a new path to further improve the performance of SOT-based magnetic devices.

## I. INTRODUCTION

Electrical manipulation of magnetization by spin orbit torque (SOT) is essential in developing long-endurance and ultrafast magnetic memory applications, such as magnetic random access memory (MRAM).[1-5] With strong spin orbit interaction, a pure charge current is converted into a spin current through spin Hall effect (SHE) or Rashba effect.[6-14] In heterostructures such as heavy metal / ferromagnet (HM / FM) bilayers, the interfacial spin accumulation exchanges its angular momentum with the FM layer, which results in a damping-like (DL) torque, as shown in Fig. 1(a). For ferromagnetic materials, this DL torque has an interface nature and it can be enhanced through increasing the interfacial spin mixing conductance ($G_r$).[15-24]

Conventionally, the critical current of SOT switching was considered to be determined by the DL torque efficiency ($\theta_{DL}^{ST}$).[25] Recently, people noticed that another component of SOT, which is called the field-like (FL) torque, also plays a central role in the ultrafast SOT dynamics. T. Taniguchi *et al.* reported that the critical current destabilizing the perpendicular magnetization by SHE depends significantly on the ratio between the FL and DL torque ($\theta_{FL}^{ST}/\theta_{DL}^{ST}$).[26] J. M. Lee *et al.* and J. Yoon *et al,* reported that a relatively large FL torque can induce anomalous switching back phenomena and damage the robustness of SOT switching.[27,28] Moreover, in magnetic racetrack memory, the velocity of SOT driven domain-wall motion was also reported to be closely related to a FL torque.[29,30] Therefore, tuning $\theta_{FL}^{ST}/\theta_{DL}^{ST}$ may serve as another approach to further improve the performance of SOT-based magnetic device.

However, as far as our knowledge, there is not yet an effective way to modulate $\theta_{FL}^{ST}/\theta_{DL}^{ST}$. In heterostructures, a FL torque is largely derived from the exchange interaction between the magnetization and the interfacial spin accumulation.[31-38] The strength of this exchange interaction can be characterized by the imaginary part of the interfacial spin mixing conductance $G_i$, which is related with certain material system. Up to now, there is no report about the experimental adjustment of $G_i$ and a related FL torque. Here, we propose to modulate a FL SOT by controlling the amount of interfacial spin accumulation, through tuning the interfacial spin transparency.[37,38] Recent study in HM/FM/capping trilayers shows that the interfacial spin transmission is not only related with the strength of the interfacial spin orbit coupling (ISOC), lattice mismatch or other interfacial conditions, but also with the spin memory loss happening apart from the interface.[23] Therefore, the effective interfacial spin mixing conductance $G_r$, and hence a FL SOT, could be tuned by modulation of the spin dephasing inside FM layer, which is related with the relative distance between the layer thickness and the spin diffusion length.

We experimentally modulate $\theta_{FL}^{ST}/\theta_{DL}^{ST}$ in W/CoFeB/MgO trilayers with perpendicular magnetic anisotropy (PMA). Different layer thicknesses and annealing temperatures are adopted to control the spin dephasing in CoFeB. By spin Hall magnetoresistance (SMR) measurement, we demonstrate that the FM layer with larger spin dephasing can unambiguously boost spin absorption and improve the effective $G_r$. $\theta_{FL}^{ST}$ and $\theta_{DL}^{ST}$ are extracted by a standard harmonic measurement.[21,39] While $\theta_{DL}^{ST}$ increases with $G_r$, $\theta_{FL}^{ST}$ turns out to decrease with $G_r$. The results suggest that a FL SOT is largely contributed by the exchange coupling between the magnetization and the interfacial spin accumulation, which decreases with larger effective $G_r$. This modulation of $\theta_{FL}^{ST}/\theta_{DL}^{ST}$ is more obvious when the spin generation efficiency is lower. Our findings suggest a new direction to further improve the performance of SOT-based magnetic devices by modulation of FL torque amplitude through engineering of the interfacial spin transparency.

## II. RESULTS AND DISCUSSIONS
### 2.1 Sample preparation and fundamental properties

The films studied are deposited on a 300 nm oxidized Si substrate under room temperature. 1h annealing at certain temperature ($T_{AN}$) is performed after deposition to enhance the PMA.[40] Three series of samples are fabricated for comparison: I W(1.5)/CoFeB($t_{CoFeB}$)/MgO(2)/Ta(1.5) $t_{CoFeB}$ = 0.9~1.5 nm with 0.1 nm step, $T_{AN}$ = 300 ℃. II W(1.5)/CoFeB(1.1)/MgO(2)/Ta(1.5), $T_{AN}$ = AD, 150, 200, 250, 300, 325, 350 ℃. III W(5)/CoFeB(1.1)/MgO(2)/Ta(1.5), $T_{AN}$ = AD, 200, 250, 300, 350, 400 ℃, where AD hereafter represents as deposited sample with no annealing. Numbers in parentheses are layer thicknesses in nanometers. The CoFeB composition is $Co_{20}Fe_{60}B_{20}$ (simplified as CoFeB in this paper). All layers are deposited by magneto sputtering under an Argon pressure of 1 mTorr. The chamber base pressure is less than $3\times10^{-8}$ Torr. A deposition rate of less than 0.02nm/s is used for tungsten layer to obtain β-phase tungsten.[6] Fig. 1(b) shows the transmission electron microscope (TEM) image of W(1.5)/CoFeB(1.1)/MgO(2)/Ta(1.5), $T_{AN}$ = 300 ℃. A standard photolithography, dry etching and lift off process is then performed to fabricate hall bar structure and electrode (Ti20/Au100). A relatively small ratio (0.25) between voltage channel and current channel width is designed to restrain shunting effect at hall bar cross area.[41]

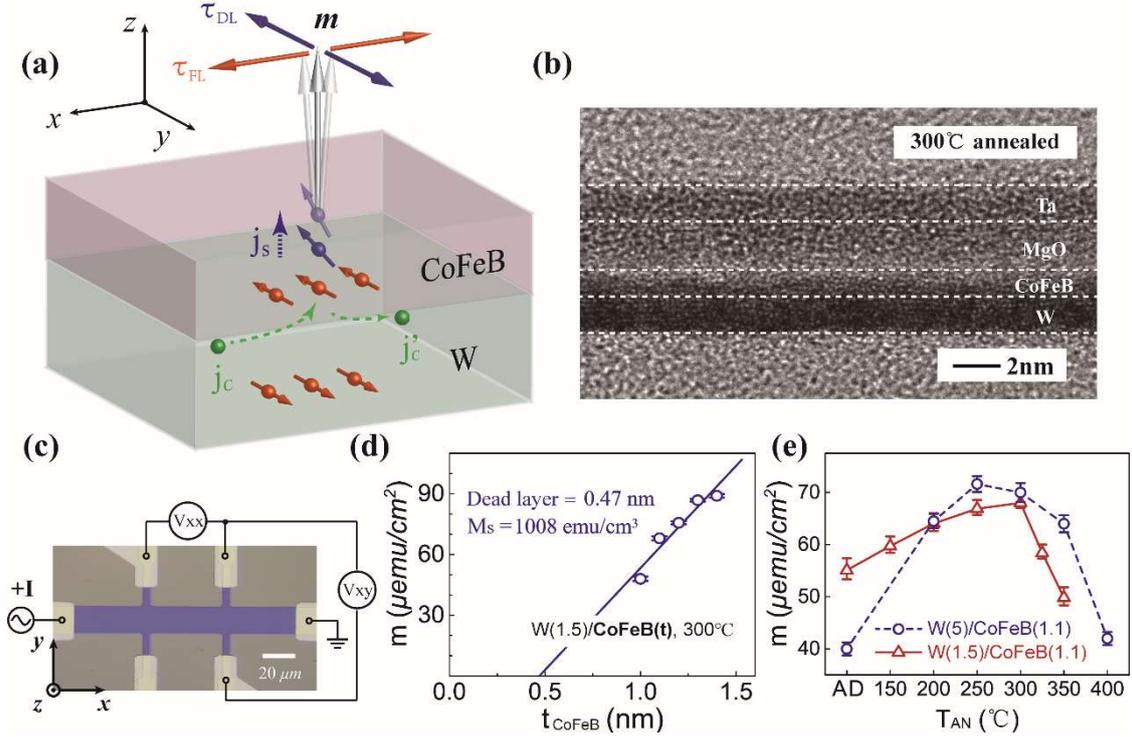

**Figure 1.** (a) Mechanism of DL and FL SOT in W/CoFeB bilayers. The red spins represent the interfacial spin accumulation, which largely contributes to a FL torque by exchange interaction. The purple spins represent the spin absorption of the CoFeB layer as a spin transfer torque (DL torque). $j_c$ and $j'_c$ represent the applied current and the current generated through ISHE by spin diffusion. (b) TEM image of W(1.5)/CoFeB(1.1)/MgO(2)/Ta(1.5), $T_{AN}$ = 300 ℃. (c) The transport measurement illustrated on an exemplary microscope image. (d) and (e) Saturation magnetization per unit area of sample series I (d), II and III(e).

Fig. 1(c) shows the transport measurement illustrated on an exemplary microscope image. The 1.5 nm Ta capping is considered as totally oxidized and its shunting effect is ignored. The resistivity of W and CoFeB are evaluated to be $\rho_W = 183\,\mu\Omega\cdot cm$ and $\rho_{CoFeB} = 180\,\mu\Omega\cdot cm$ (See Supporting information for more details about resistivity measurement). In sample series II, the resistivity of the multilayers keeps approximately the same during annealing. The results indicate that the W layer has a dominant β phase and a constant $\theta_{SH}$ before and after annealing.[6] The saturation magnetization m (per unit area) is measured by vibrating sample magnetometer (VSM). Fig. 1(d) shows m of sample series I. m increases with CoFeB thickness $t_{CoFeB}$. In sample series II and III, m increases with $T_{AN}$ under ~300 °C, as shown in Fig. 1(e). Higher than ~300 °C, there is an over annealing process that leads to the decrease of magnetization.[40]

**2.2 Characterization of interfacial spin transmission by spin Hall magnetoresistance**

Conventionally, the interfacial spin mixing conductance was frequently characterized by gigahertz ferromagnetic resonance (FMR) measurement, where the damping factor α is utilized as an indicator of $G_r$.[3,15] However, recently, ref.42 reported that the two-magnon scattering (TMS) caused by ISOC will lead to an overestimation of α. Since the ISOC may change upon annealing, the FMR detection may not correctly characterize the $G_r$ variation in our annealing experiment.

We utilize the spin Hall magnetoresistance (SMR) measurement to characterize the effective $G_r$.[37,38,43-45] As shown in Fig. 1(a), in W/CoFeB bilayers, the spin reflection from the interface generates a charge current $\vec{j_c'}$ with the same direction to $\vec{j_c}$ by inverse spin Hall effect (ISHE), which increases the W conductivity. When $\vec{m}$ is perpendicular to the spin polarization $\vec{\sigma}$, a large part of the interfacial spin accumulation is absorbed by $\vec{m}$ as a spin transfer torque, which minimizes $\vec{j_c'}$ and W conductivity. When $\vec{m}$ is parallel or antiparallel with $\vec{\sigma}$, there is only a slight reduction of interfacial spin accumulation through longitudinal spin absorption by bulk FM, which maximizes $\vec{j_c'}$ and W conductivity. Therefore, there is a magnetoresistance (MR) in W as a function of the relative angle between $\vec{m}$ and $\vec{\sigma}$. This so called SMR can be utilized to characterize the effective $G_r$ (See Supporting information for formula derivation).[38]

We measure the spin Hall magnetoresistance with 100 μA DC current under an external field of 6 T rotating in y-z plane. Since the magnetization is always perpendicular with the applied current, the anisotropy magnetoresistance (AMR) in bulk FM layer, which depends on the relative angle between the magnetization and the applied current, is excluded.[43,44] Ref. 46 shows that a FM single layer has another MR among y-z plane called anomalous Hall magnetoresistance (AHMR). We determine the AHMR of CoFeB layer to be 1~2 orders smaller compared to SMR (see Supporting information). Therefore, the CoFeB resistance is treated as a constant. $R_W^{XX}(\beta)$ can be extracted from overall longitudinal resistance $R^{XX}(\beta)$ by:

$$(R^{XX}(\beta))^{-1} = \left(R_W^{XX}(\beta)\right)^{-1} + \left(R_{CoFeB}^{XX}\right)^{-1} \quad (1)$$

, where $R_{CoFeB}^{XX} = \rho_{CoFeB} \, l/(w \cdot t_{CoFeB})$, l is the longitudinal length between two abreast voltage channels.

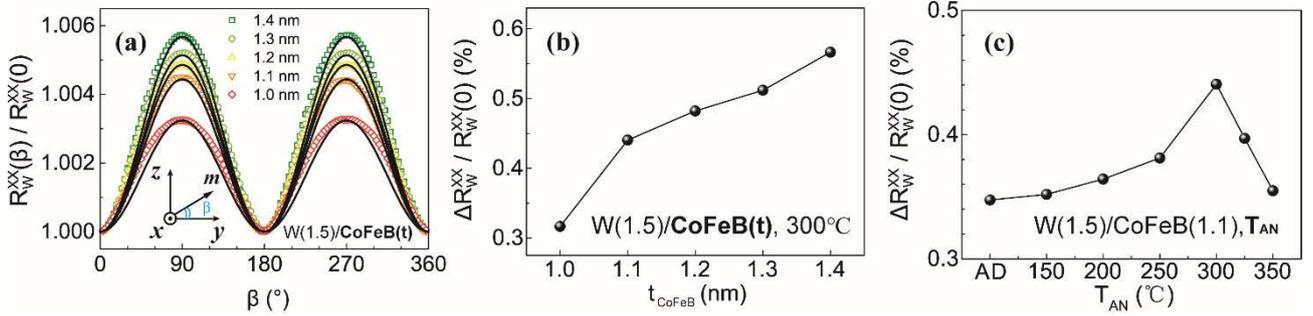

**Figure 2.** Characterization of interfacial spin transmission by SMR measurement. The SMR ratio reveals the effective interfacial spin mixing conductance $G_r$. (a) SMR measurement of sample series I. The inset shows the schematic of magnetization rotation. (b) and (c) Summarized SMR ratio of sample series I (b) and II (c).

Fig. 2(a) shows the SMR measurement of sample series I. As summarized in Fig. 2(b), the normalized SMR amplitude increases with $t_{CoFeB}$, which indicates that the $G_r$ also increases with $t_{CoFeB}$. We attribute this phenomenon to the following reason. Within the spin diffusion length, larger thickness can improve the exchange of angular momentum inside the CoFeB layer, and thus improves the spin penetration. Fig. 2(c) shows the SMR amplitude as a function of $T_{AN}$ in sample series II. The spin transmission increases with $T_{AN}$ upon proper annealing and decreases with $T_{AN}$ upon an over annealing process beyond ~300 °C. By comparing Fig. 2(c) with Fig. 1(e), we can find that $G_r$ is positively related with $M_S$. This indicate that stronger magnetization reduces the spin diffusion length of CoFeB, which improves the momentum exchange and the effective $G_r$.

## 2.3 Harmonic characterization of $\theta_{FL}^{ST}$ and $\theta_{DL}^{ST}$

To verify the modulation of FL torque, we perform a standard harmonic characterization considering planar Hall effect correction.[21,39] Fig. 3(a) and Fig. 3(b) shows the exemplary harmonic measurement results of W(1.5)/CoFeB(1.1)/MgO(2)/Ta(1.5), $T_{AN}$ = 300 °C. See Supporting information for details about extraction of transverse (longitudinal) effective fields $\Delta H_{y(x)}$. We extract the intrinsic FL (DL) SOT efficiency $\theta^{ST}_{FL(DL)}$ from the $\Delta H_{y(x)}$ by:[47]

$$\theta^{ST}_{FL(DL)} = \frac{2|e|M_s t_{CoFeB}}{\hbar} \times \left|\frac{\Delta H_{y(x)}}{J_W}\right| \quad (2)$$

Fig. 3(c) summarizes the FL and DL $\theta^{ST}$ of sample series I. As $t_{CoFeB}$ increases from 1.0 nm to 1.4 nm, the $\theta^{ST}_{DL}$ increases slightly from 0.063±0.004 to 0.066±0.002, while the $\theta^{ST}_{FL}$ decreases rapidly from 0.146±0.005 to 0.060±0.002. The trend of $\theta^{ST}_{FL}/\theta^{ST}_{DL}$ is consistent with previous study by *J. Kim et al.*[21] Fig. 3(d) shows the $\theta^{ST}$ as a function of $T_{AN}$ in W(1.5)/CoFeB(1.1)/MgO(2)/Ta(1.5). The $\theta^{ST}_{FL}$ decreases with $T_{AN}$ upon proper annealing and increase again during an over annealing process higher than ~300°C. Meanwhile, the $\theta^{ST}_{DL}$ changes inversely with $\theta^{ST}_{FL}$. By comparison of Fig. 2 and Fig. 3, we find that a FL torque can be effectively diminished by larger interfacial spin transparency.

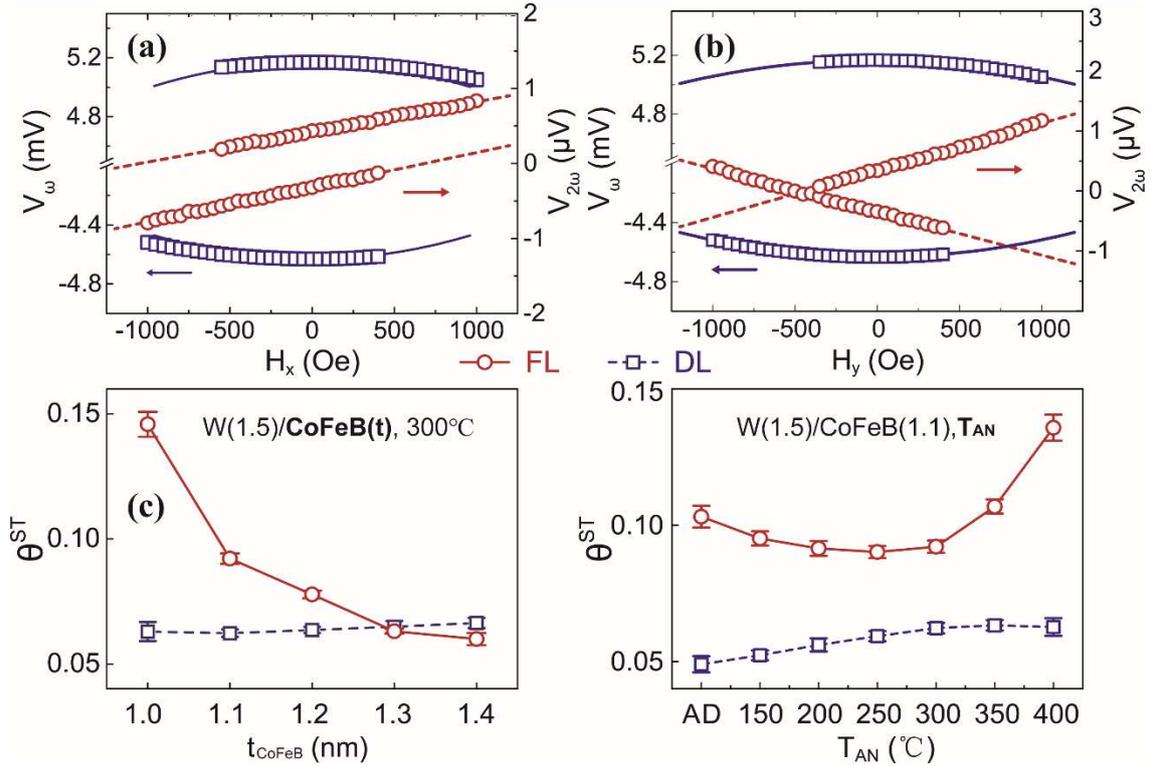

**Figure 3.** (a) and (b) First and second harmonic Hall voltage ($V_\omega$ and $V_{2\omega}$) measured on W(1.5)/CoFeB(1.1)/MgO(2)/Ta(1.5), $T_{AN}$ = 300 °C with an external magnetic field sweeping along x (a) or y (b) axis. (c) and (d) FL torque and DL torque efficiency of sample series I (c) and II (d).

Next, we discuss how does larger effective $G_r$ reduce a FL spin torque. When we neglect the transverse spin diffusion into bulk FM layer, spin orbit torque can be expressed as:[37]

$$\tau = \frac{\hbar}{2e^2}[G_r \hat{m} \times (\hat{m} \times \vec{\mu_s}) + G_i \hat{m} \times \vec{\mu_s}] \quad (3)$$

, where $\vec{\mu_s}$ is spin accumulation. In Eq. 3, the first term in bracket has a similar form of the damping term in the Landau-Lifshitz-Gilbert equation and it's so called damping-like torque. Its strength is determined by $\vec{\mu_s}$ and $G_r$, where $\vec{\mu_s}$ depends on both $\theta_{SH}$ and $G_r$. Comparably, the second term in bracket is called the field-like torque, whose amplitude is decided by $G_i$ and $\vec{\mu_s}$. $G_i$ is the imaginary part of spin mixing conductance and it can be interpreted as an effective exchange field. The amplitudes of DL and FL torques are both related to the integral of $\vec{\mu_s}$ weighted by the sine of the angle between $\vec{\mu_s}$ and $\vec{m}$. The interfacial spin accumulation doesn't transfer its momentum to the magnetization, and it exerts only a FL torque through exchange interaction. Given the fact that injected spin current is tilted progressively toward the magnetization, it gives smaller FL torque compared to the interfacial spin accumulation. Therefore, the enhancement of effective $G_r$ will reduce a FL torque.

To compare the strength of this modulation under different efficiencies of spin generation, we further measure the FL torque in sample series III with a larger W thickness of 5 nm, as shown in Fig. 4(a). Fig. 4(b) shows the comparison of $\theta_{FL}^{ST}/\theta_{DL}^{ST}$ between sample series II and III. In sample series II with a thinner W layer, $\theta_{FL}^{ST}/\theta_{DL}^{ST}$ shows stronger dependence on $T_{AN}$. This can be interpreted by the smaller amount of spin generation in thinner W layer. With fewer initial interfacial spin accumulation, the enhancement of spin absorption upon annealing takes a larger proportion. As a result, for structures with less spin generation, the modulation of $\theta_{FL}^{ST}/\theta_{DL}^{ST}$ by annealing is more effective.

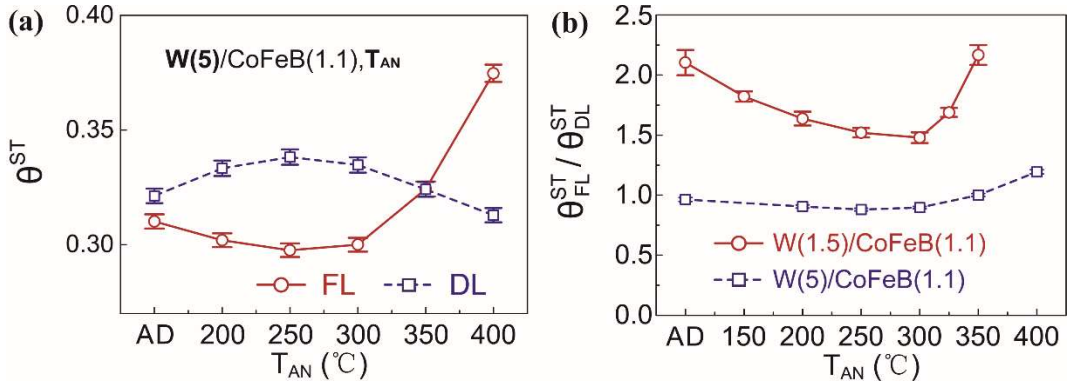

**Figure 4.** (a) FL torque and DL torque efficiency of sample series III. (b) $\theta_{FL}^{ST}/\theta_{DL}^{ST}$ in sample series II and III as a function of $T_{AN}$.

The performance of SOT switching can be enhanced by the reduction of $\theta_{FL}^{ST}/\theta_{DL}^{ST}$. Ref. 28 experimentally showed that when the FL has an opposite sign with DL, which is also the case in W/CoFeB/MgO trilayers, a large FL torque will induce anomalous switching back phenomena in nanosecond-timescale, which decreases the switching probability. One of our recent theoretical work also demonstrated that, with the same sign relation, a larger $\theta_{FL}^{ST}$ will reduce the robustness of SOT switching.[48] Although, in our experiment, larger total magnetization moment is needed to reduce $\theta_{FL}^{ST}$ and it will increase the critical current of SOT switching, it also increases the thermal stability and enables the scaling-down of memory devices. For SOT-MRAM, there is a trade-off between the power consumption and other properties, such as thermal stability, switching speed and robustness.

At the end of our discussion, we are able to depict the mechanism of FL and DL SOT. As shown in Fig. 1(a), a horizontal charge current generates a spin current by SHE or Rashba effect. A part of the spin current diffuses into the FM layer and exerts a DL torque toward $\vec{\sigma}$ by spin transfer torque, and a FL torque around $\vec{\sigma}$ by exchange interaction. Additionally, the spin accumulation at the interface contributes solely to a FL torque. This significant origin of a FL torque could be reduced by improving the interfacial spin transmission. The spin current reflection caused by interfacial spin accumulation converts back into a charge current by ISHE, which could be detected by SMR measurement.

## III. CONCLUSION

In summary, without the need to change material system in order to tune $G_i$, we achieve modulation of $\theta_{FL}^{ST}/\theta_{DL}^{ST}$ by control of $G_r$ in W/CoFeB/MgO trilayers with PMA. With the help of SMR measurement, we find that thicker or properly annealed CoFeB with larger spin dephasing can obviously boost the absorption of spin current and increase the interfacial spin transmission. $\theta_{FL}^{ST}$ is found to be reduced by larger $G_r$. Moreover, this modulation of FL torque is found to be even more effective in structures with smaller spin generation. The results indicate that the FL SOT is strongly contributed by the exchange interaction between the magnetization and interfacial spin accumulation, which is reduced by larger interfacial spin transparency. We suggest a new routine to further enhance the performance of SOT-based magnetic device by modulating the relative amplitude of FL torque through control of interfacial spin transparency.

**ACKNOWLEDGMENTS**

We thank Prof. Luqiao Liu for fruitful discussion. The authors gratefully acknowledge the National Key R&D Program of China (No. 2018YFB0407602), National Natural Science Foundation of China (Grant No. 61627813, 61571023), the Beihang Hefei Innovation Research Institute Project BHKX-17-06, the International Collaboration Project B16001, the VR innovation platform from Qingdao Science and Technology Commission and the National Key Technology Program of China 2017ZX01032101 for their financial support of this work.

**Supporting information:**

Resistivity measurement of sample series I and II. Formula derivation of SMR measurement. Anomalous Hall magnetoresistance in single CoFeB layer. Details of harmonic measurement.

# Supporting information

## S1. Resistivity measurement of sample series I and II

We measure the longitudinal conductance ($\sigma_{XX}$) of sample series I and II. As illustrated in Fig. S1(c), $\sigma_{XX}$ is extracted by $\sigma_{XX} = I/V_{XX}$. The resistivities of W and CoFeB are evaluated from the slope and intercept of the linear fitting in sample series I. $\rho_W = 183\ \mu\Omega\ \text{cm}$ and $\rho_{CoFeB} = 180\ \mu\Omega\ \text{cm}$. As shown in Fig. S1(b), the resistance keeps approximately a constant during annealing.

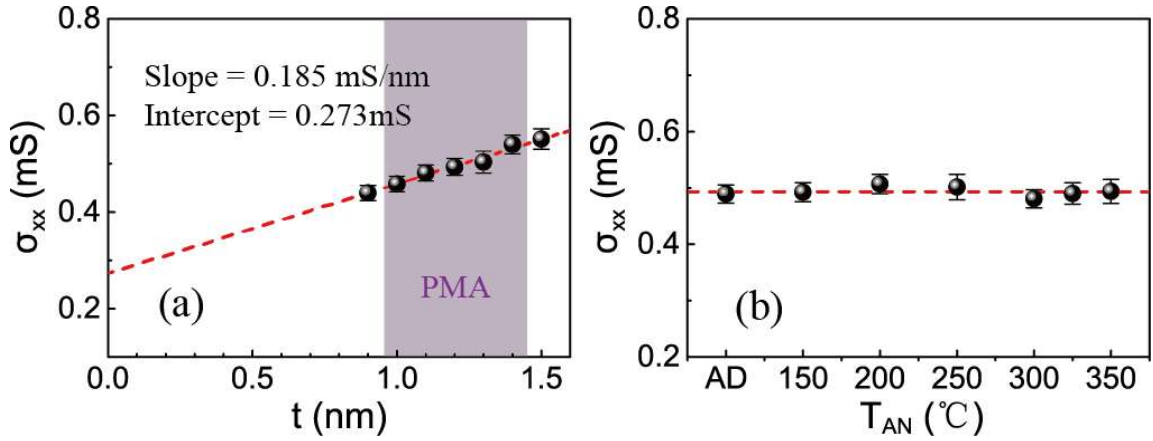

**Figure S1.** Conductivity of sample series I (a) and II (b). Every data point is an average of 3~7 hall bar structures.

## S2. Formula derivation of SMR measurement

Considering the longitudinal spin absorption of the FM layer, the SMR ratio $\Delta R_W^{XX}/R_W^{XX}(0)$ can be expressed by following equation (shunting effect term is removed by considering only the HM layer resistance):[S1,S2]

$$\frac{\Delta R_W^{XX}(\beta)}{R_W^{XX}(z)} = -\theta_{SH}^2 \frac{\lambda_{HM}}{t_{HM}} \tanh^2(t_{HM}/2\lambda_{HM})$$

$$\times \left[ \frac{g_R}{1 + g_R \coth(t_{HM}/\lambda_{HM})} - \frac{g_F}{1 + g_F \coth(t_{HM}/\lambda_{HM})} \right]$$

$$g_R \equiv 2\rho_{HM}\lambda_{HM} G_r, \quad g_F \equiv \frac{(1-P^2)\rho_{HM}\lambda_{HM}}{\rho_{FM}\lambda_{FM}\coth(t_{FM}/\lambda_{FM})} \quad (S1)$$

, where $\lambda_{HM}$, $t_{HM}$ and $\rho_{HM}$ are the spin diffusion length, thickness and resistivity of the HM layer. P, $\lambda_{FM}$, $t_{FM}$ and $\rho_{FM}$ are the spin polarization, spin diffusion length, thickness and resistivity of the FM layer. $R_W^{XX}(z)$ is the W resistance when m points ±z direction. The first term in bracket is the major source of SMR caused by variation of interfacial spin transfer torque. Its strength is related to both the efficiency of current-spin conversion ($\theta_{SH}$) and spin mixing conductance ($G_r$). Since $\theta_{SH}$ of each sample series keeps a constant, the SMR ratio dominantly reflects effective $G_r$ in our experiment. The second term in bracket represents a slight reduction of SMR caused by longitudinal spin current absorption, whose amplitude increases with $t_{FM}$.[S2] Since $\Delta R_W^{XX}/R_W^{XX}(0)$ increases monotonously with $t_{CoFeB}$ in sample series I, we verify that this slight reduction caused by longitudinal spin current absorption is not the dominant source of SMR variation in our experiment. The effective interfacial spin transparency $T_{int}$, which is the proportion of spin transfer torque in total spin generation $\theta_{DL}^{ST}/\theta_{SH}$, has a positive correlation with the effective $G_r$,

$$T_{int} = \text{Re}\left\{ \frac{2G_{\uparrow\downarrow}\tanh(t_{HM}/\lambda_{HM})}{\sigma_{HM}/\lambda_{HM} + 2G_{\uparrow\downarrow}\coth(t_{HM}/\lambda_{HM})} \right\} \quad (S2)$$

, where $\sigma_{HM}$ is the conductivity of the HM layer.

We notice that in Ref. S3, V. L. Grigoryan et al. predicted that the Rashba effect also contributes to a SMR. Since the strength of ISOC and Rashba effect may vary during annealing,[S4] our current explanation of SMR data may be disproved. However, we point out that, if the strengthen or weakening of Rashba effect is the dominant source of SMR variation in our annealing experiment, it will lead to the improvement or reduction of the efficiency of current-spin conversion ($\theta_{SH}$). As a result, both the DL and FL torque will increase or decrease simultaneously, which is not the case in our experiment.

**S3. Anomalous Hall magnetoresistance in single CoFeB layer**

Y. Yang et al. reported that there exists a novel kind of magnetoresistance called anomalous Hall magnetoresistance (AHMR) in FM single layer among y-z plane.[S5] We tested longitudinal AHMR in single CoFeB layers of 4 nm and 8 nm under 100 μA DC current and 3 T external field rotating inside y-z plane. As shown in Fig. S2, the AHMR of 8 nm CoFeB is ~ 0.0145 % and AHMR of 4 nm CoFeB is ~ 0.0048 %, which is negligible compared to the SMR observed in W/CoFeB/MgO.

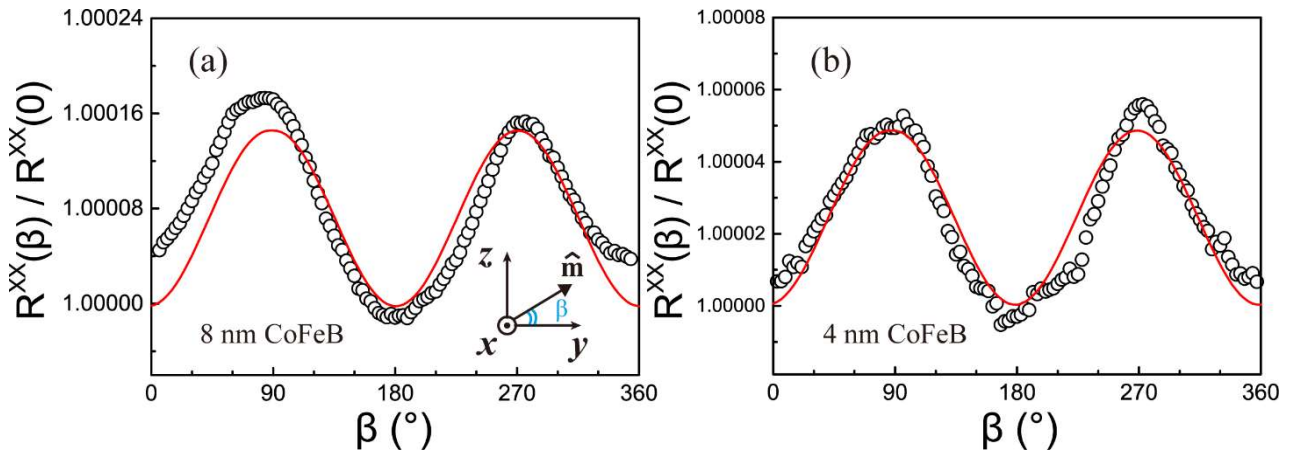

**Figure S2.** Longitudinal AHMR measured in single CoFeB layer of 8 nm (a) and 4 nm (b). The inset shows the rotation of magnetization. The red lines show the sinusoidal fitting.

## S4. Details of harmonic measurement

Longitudinal (transverse) SOT effective field $\Delta H_x$ ($\Delta H_y$) can be detected by using a standard harmonic measurement.[S6,S7] We first measure the $R_{PHE}$ under 100 μA DC currect while a 6 T external field rotating inside the *x-y* plane. Fig. S3(a) shows the $R_{PHE}$ and sinusoidal fitting measured in W(1.5)/CoFeB(1.1)/MgO(2)/Ta(1.5), $T_{AN}$ = 300 °C.

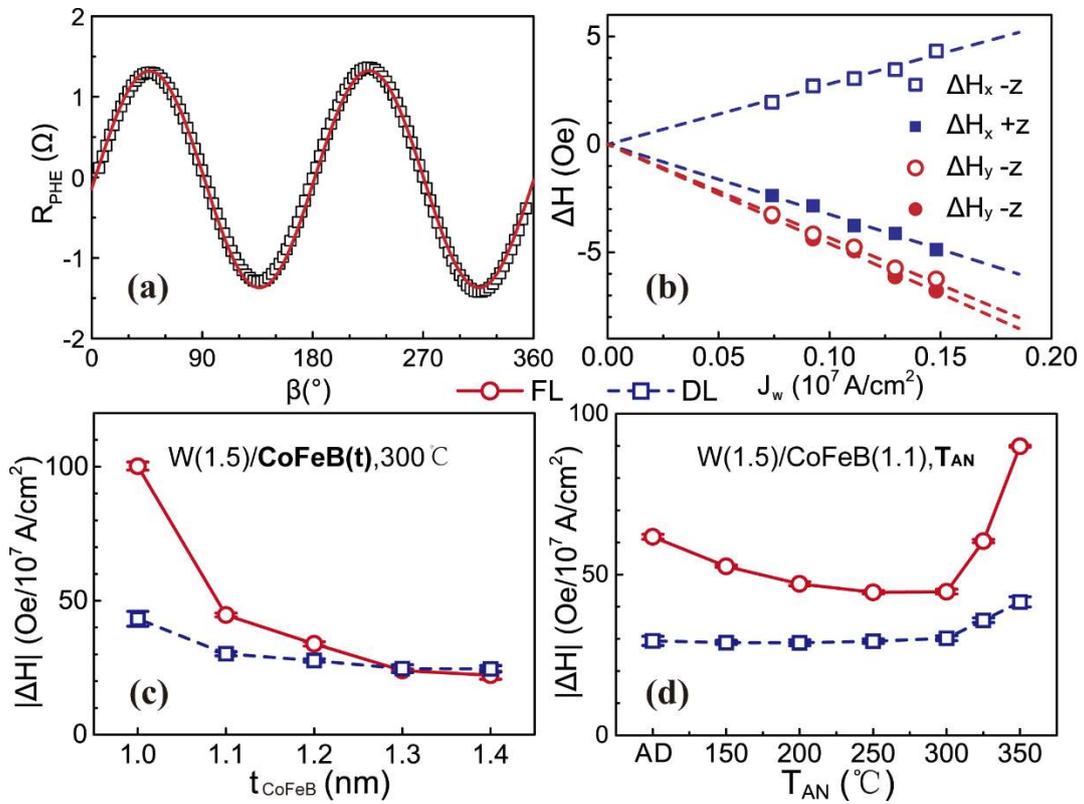

**Figure S3.** (a) Planar Hall resistance ($R_{PHE}$) of W(1.5)/CoFeB(1.1)/MgO(2)/Ta(1.5), $T_{AN}$ = 300 °C. (b) Transverse and longitudinal effective fields as a function of current density of W. (c) and (d) FL and DL effective fields of sample series I and II. Error bars come from the linear fitting of ΔH as a function of current density.

$\Delta H_x$ ($\Delta H_y$) is extracted from the parabolic and linear fit of 1st ($V_\omega$) and 2nd ($V_{2\omega}$) harmonic voltage using the following equation considering planar hall effect (PHE) correction:

$$\Delta H_{x(y)} = -\frac{(B_{x(y)} \pm 2rB_{y(x)})}{1-4r^2}, \quad B_{x(y)} \equiv 2\left(\frac{\partial V_{2\omega}}{\partial H_{x(y)}}\right)/\left(\frac{\partial^2 V_\omega}{\partial H_{x(y)}^2}\right) \quad (S3)$$

, where ± stands for ±z magnetization directions, r is the ratio between PHE resistance ($R_{PHE}$) and $R_{AHE}$, $H_x$ ($H_y$) is an external magnetic field sweeping along x (y) direction. The frequency of the adopted AC current is 133.33Hz. We tilted the sample by 2~4° to give a small out of plane external field in order to prevent multi-domain formation. The average value of the effective fields extracted with the magnetization pointing ±z direction with a small tilt angle can be a good approximation of the actual value.[S8] The current density in W layer is calculated by

$$J_W = J \cdot \frac{\rho_{CoFeB}/t_{CoFeB}}{\rho_{CoFeB}/t_{CoFeB} + \rho_W/t_W} \quad (S4)$$

, where J and $J_W$ are the current density flowing through the full stack and the W layer respectively, and $t_W$ is the W thickness. Fig. S3(b) shows the obtained SOT effective field under different current density. $\Delta H_x$ (DL) and $\Delta H_y$ (FL) (per $10^7 A/cm^2$) are extracted from the linear fitting of the effective field. Fig. S3(c) and 3(d) summarize the SOT effective field of sample series I and II. We can find that $\Delta H_x$ and $\Delta H_y$ are both reduced by larger magnetization. With the same SOT efficiency, SOT effective field is inversely proportional to the magnetization.